\documentclass[twocolumn,showpacs,10pt,prb,floatfix,amssymb]{revtex4}

\usepackage{graphicx}
\newcommand{\Idet}{I_\mathrm{det}}
\newcommand{\Iinj}{I_\mathrm{inj}}
\newcommand{\Vsd}{V_\mathrm{sd}}
\newcommand{\Vp}{V_\mathrm{p}}

\begin{document}

\newlength{\plotwidth}
\setlength{\plotwidth}{8.5cm}

\title{Ballistic electron spectroscopy}

\author{F.~Hohls}
\author{M.~Pepper}
\author{J.~P.~Griffiths}
\author{G.~A.~C.~Jones}
\author{D.~A.~Ritchie}
\affiliation{Cavendish Laboratory, University of Cambridge, J J Thomson Avenue,
Cambridge CB3 0HE, UK}

\date{\today}

\begin{abstract}
We demonstrate the feasibility of ballistic electron spectroscopy as a new tool
for mesoscopic physics. A quantum dot is utilised as an energy-selective detector of
non-equilibrium ballistic electrons injected into a two-dimensional electron
system. In this paper we use a second quantum dot as the electron injector to
evaluate the scheme. We propose an application in the study of interacting 1D
and 0D systems. % changed 1d/0d to 1D/0D
\end{abstract}
\pacs{73.23.Ad, 73.23.Hk} \maketitle

%%%%%%%%%%%%%%%%%%%%%%%%%%%%%%%%%%%%%%%%%%%%%%%%%%%%%%%%%%%%%%%%%%%%%
% Introduction
%%%%%%%%%%%%%%%%%%%%%%%%%%%%%%%%%%%%%%%%%%%%%%%%%%%%%%%%%%%%%%%%%%%%%

Quantum dots and wires are coming to an age of application and are used as
building blocks of devices with increasing complexity.
\cite{Elzerman2004,Petta2005} This sets high standards in the characterization
and understanding of the individual elements. Non-equilibrium transport
measurement, also named transport spectroscopy, is amongst the most commonly %%
used tool to reveal characteristics of quantum wires \cite{Patel1991} and
quantum dots.\cite{Kouwenhoven1997rev} For one-dimensional systems, the
sub-band spacing can be determined; for quantum dots, level spacing and
tunnelling rates can be extracted.  However, if there is strong interaction
between electrons in these systems, the internal structure cannot be assumed to
be independent of the applied bias voltage.

We propose instead to directly measure the energy distribution of
non-equilibrium ballistic electrons leaving %%
the device under study, allowing us to draw conclusions concerning the internal
energetic structure without altering it with our measurement technique.
Injection and detection of non-equilibrium electrons has been demonstrated for
MBE grown and Schottky gate induced tunnelling barriers
\cite{Heiblum1985,Heiblum1989,Mathews1990} and also using quantum point
contacts.\cite{Yacoby1991,Williamson1990,Dzurak1992} However, these techniques
do not offer true energy selectivity as electrons at all energies beyond a
threshold can pass the detector. {\sl Here} we use the controllable energy
selectivity of resonant transport through a quantum dot as a spectrometer of
non-equilibrium ballistic electrons which allows us to assess the complete
energy distribution.

%%%%%%%%%%%%%%%%%%%%%%%%%%%%%%%%%%%%%%%%%%%%%%%%%%%%%%%%%%%%%%%%%%%%%
%    Paper bulk
%%%%%%%%%%%%%%%%%%%%%%%%%%%%%%%%%%%%%%%%%%%%%%%%%%%%%%%%%%%%%%%%%%%%%

\begin{figure}[b]
            \vspace{-3mm}
    \begin{center}
            \includegraphics[width=0.95\plotwidth]{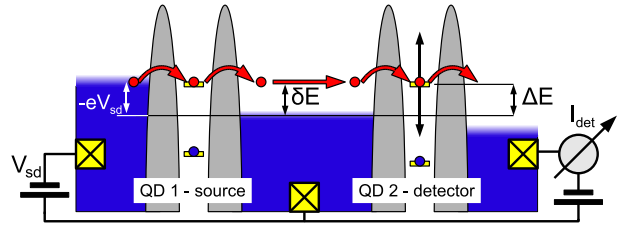}
    \end{center}
           \vspace{-3mm}
    \caption{%
Schematic of ballistic electron spectroscopy: The left quantum dot (QD1)
injects non-equilibrium ballistic electrons into the middle 2DES region. Those
electrons that travel to the detector quantum dot (QD2) can only tunnel to the
right lead if their energy matches one of the detector dot states, the energy
of which can be tuned by a gate. Recording the current $I$ as function of the
dot energy allows measurement of the spectrum of the incident electrons.
        }
    \label{fig1}
\end{figure}

In this work we test the feasibility of our ballistic electron spectroscopy
scheme using a second quantum dot for a well-defined injection of
non-equilibrium electrons. The schematic of our device is shown in
Fig.~\ref{fig1}. Resonant tunnelling through the left quantum dot injects
ballistic electrons into the middle two-dimensional electron system (2DES)
region. Those electrons that travel ballistically to the detector dot and whose
excess energy $\delta E$ matches the resonance condition ($\delta E=\Delta E$)
can tunnel to its drain and are detected as a current. We change the resonant
energy of the detector $\Delta E$ to scan the complete spectrum of the incident
ballistic electrons. A magnetic field $B$ perpendicular to the plane of the
2DES allows us to differentiate the ballistic portion of the signal.

We use standard Schottky metal gates\cite{Thornton1986} to define our device in
a high mobility 2DES in a GaAs/AlGaAs heterostrucure. The 2DES lies 90~nm
beneath the surface and has a density $n_e = 1.7\!\times\!10^{15}$~m$^{-2}$ and
a mobility  $\mu_e = 170$~m$^2$/Vs. The transport mean free path, $l_e$, is of
order $12\,\mu$m, exceeding the dimensions of the device. All experiments were
performed with an electronic temperature $T\lesssim 0.2$~K. For this
temperature and for the excess energies of our ballistic electrons, $\delta E
\leq 0.3$~meV, we expect the dominant relaxation process to be
electron-electron scattering, with a scattering length $l_{e-e} \geq 10
\,\mu$m.\cite{Giuliani1982,Yacoby1991}

\begin{figure}[tb]
    \begin{center}
            \includegraphics[width=\plotwidth]{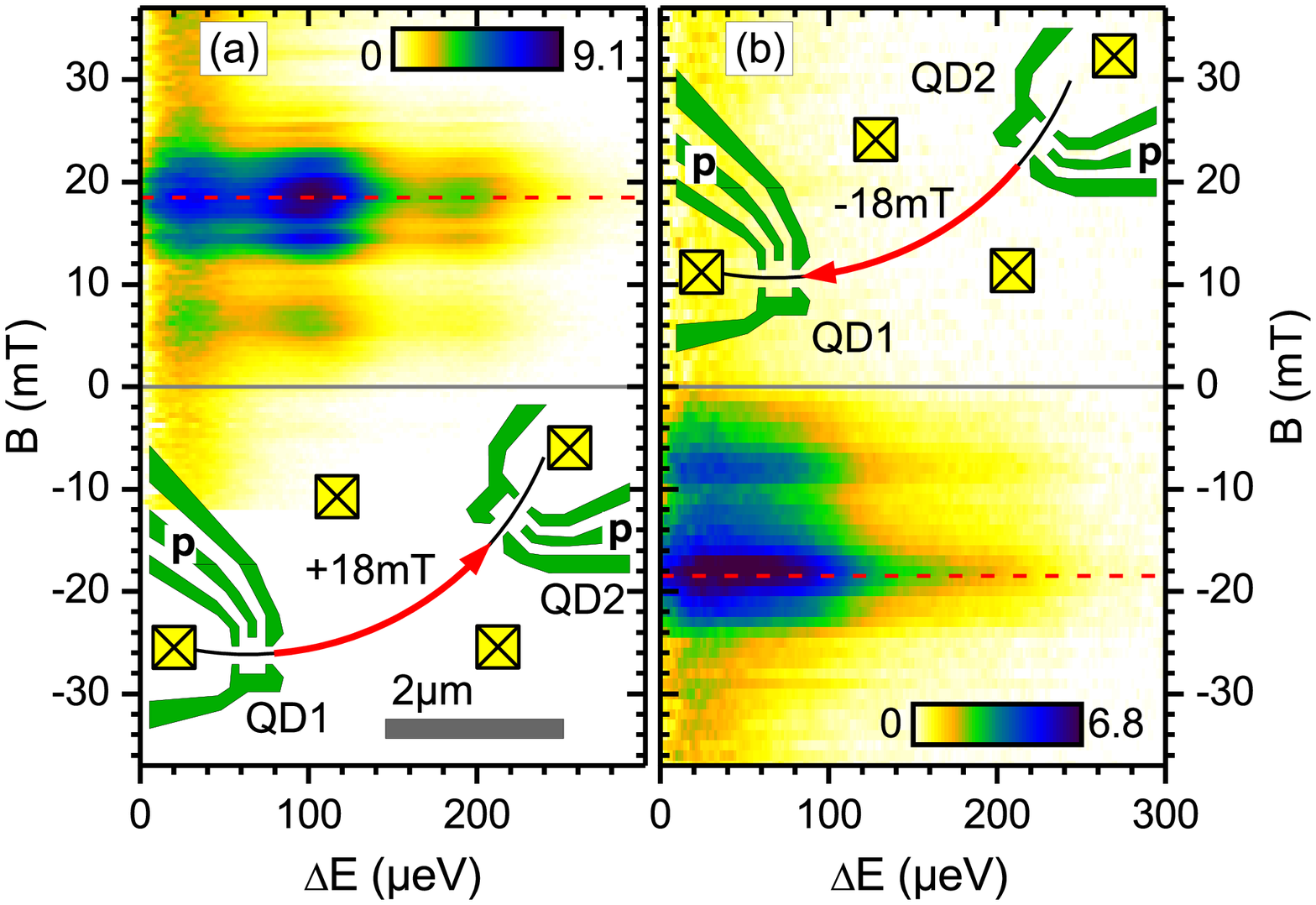}
    \end{center}
            \vspace{-4mm}
    \caption{%
Current $\Idet\,$[pA] measured by the detector dot as a function of detector
energy $\Delta E$ and magnetic field $B$ for 200$\,\mu$eV injector energy.
Panel (a) shows injection by QD1 (injected current $\Iinj\approx280\,$pA) and
detection by QD2, panel (b) shows the result of swapping injection and
detection quantum dots ($\Iinj\approx210\,$pA). The dashed lines at
$B=\pm18.5\,$mT mark the positions of maximum signal which reverses in $B$ for
the two configurations. {\it Insets:} Layout of the device; 'p' marks the
plunger gate used to tune the dot energy. The crossed squares indicate ohmic
contacts. Injected ballistic electrons travel along a curved path,
$r_c=65\,\mu$m$/B\,$[mT]. The paths shown have radius 3.6$\,\mu$m corresponding
to $\pm18\,$mT.
        }
    \label{fig2}
               \vspace{-2mm}
\end{figure}

The gate layout is displayed as inset in Fig.~\ref{fig2}. Two quantum dots of
lithographic size $300\,$nm are separated by about $3\,\mu$m. The symmetry axis
of the quantum dots are set at an angle to each other to avoid direct
backscattering. Ohmic contacts allow independent control of the chemical
potentials of all regions. With the exception of Fig.~\ref{fig2}b we use the
left quantum dot (QD1) to inject non-equilibrium ballistic electrons into the
middle region. We bias the source with the sum of a negative DC-bias $\Vsd$ and
a square-wave AC excitation ($\pm 50\,\mu$V).
 The injector level of the dot is set to
resonance with the mean chemical potential of the source lead as depicted in
Fig.~\ref{fig1}. Thus we inject ballistic electrons with an excess energy of
$\delta E \approx -e\Vsd$ into the middle region. The AC excitation modulates
the flux of the ballistic electrons (modulation depth $\sim$~50\%) and thereby
allows for a more sensitive lock-in detection scheme. Throughout the paper we
only state the AC currents in response to this modulation.

The ballistic electron spectrometer is formed by the detector quantum dot (QD2
except for Fig.~\ref{fig2}b): its resonant transmission energy $\Delta E$ is
tuned by a plunger gate and the transmitted current is measured by a lock-in
technique at the injector modulation frequency. A small bias voltage ($+
100\,\mu$V) across the detector (see Fig.~\ref{fig1}) improves the ballistic
electron signal at small energies $\Delta E$. The $\Delta E = 0$ point is
determined for each spectrometer sweep using the current response to a small AC
voltage ($ 5\,\mu$V) applied across the detector dot. The conversion factors
from plunger gate voltage $\Vp$ to energy $\Delta E$ were determined from bias
spectroscopy measurements of the individual dots.

We apply a small perpendicular magnetic field which forces the ballistic
electrons into a cyclotron orbit with radius $r_c=65\,\mu$m$/B\,$[mT]. The
device design ensures that the injected ballistic electrons can reach the
detector only for a non-zero magnetic field that gives the right curvature of
the ballistic path. This allows direct ballistic electrons to be distinguished
from scattered ones.

Fig.~\ref{fig2} shows the AC-detector current as a function of spectrometer
energy $\Delta E$ and magnetic field for an injection energy of $\delta E
\approx 200\,\mu$eV. We first consider the magnetic field dependence. In
Fig.~\ref{fig2}a we observe a strong signal around $B=+18.5\,$mT. The
corresponding radius of 3.6$\,\mu$m matches the direct path between the dots as
shown in the inset. We observe an additional maximum at 6~mT which is thought
to be related to ballistic electrons undergoing a single scattering event with
an impurity in the region between the quantum dots. The sub-structure of the
ballistic peak itself possibly reflects the angular distribution of the
injection.\cite{Oowaki1993} We do not observe any strong signal at larger
positive or at any negative magnetic fields. As an additional test of the
ballistic nature of our signal we swapped the roles of source and detector
dots. The result is shown in Fig.~\ref{fig2}b and reveals a signal distribution
which is nearly a mirror image of Fig.~\ref{fig2}a with respect to $B=0$. The
main signal of direct ballistic electrons has swapped to $B=-18.5\,$mT as the
ballistic path now has opposite curvature.
%Only the additional maxima are changed as the scattered electron
%paths are not simply reversible.
Thus the magnetic field dependence proves convincingly the direct ballistic
nature of the signals around $\pm 18\,$mT.

\begin{figure}[tb]
    \begin{center}
            \includegraphics[width=0.97\plotwidth]{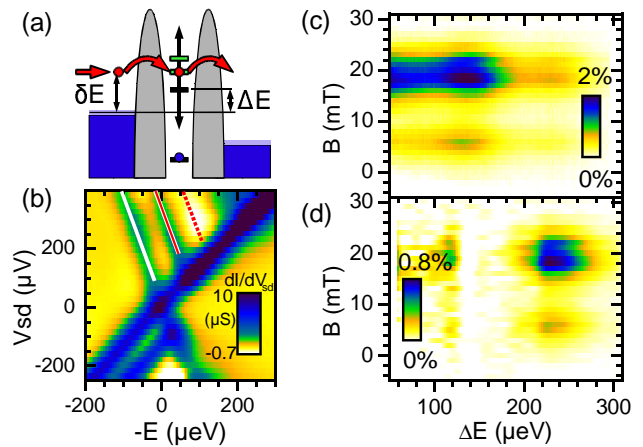}
    \end{center}
            \vspace{-4mm}
    \caption{%
a) Schematic of detected signal due to tunnelling via excited states causing
replica signal at apparent lower energies. b) Bias characterization of the
detector dot reveals excited states. %($G=dI/d\Vsd$).
The white line marks the ground state resonance with the source, the red lines
mark the resonances due to the excited states discussed in the text. c)
Normalized detector signal $\Idet/\Iinj$ as measured (injection by QD1 set to
$\approx 230\,\mu$eV, detection by QD2). d) Deconvolved signal when considering
two excited states.
        }
    \label{fig3}
            \vspace{-2mm}
\end{figure}

We now turn to the energy dependence. We inject ballistic electrons with an
excess energy of $\delta E \approx 200\,\mu$eV, and indeed we observe in
Fig.~\ref{fig2}a a peak at $\Delta E \approx 200\,\mu$eV. Thus our spectrometer
detects ballistic electrons at the correct energy. Additionally
Fig.~\ref{fig2}a reveals a large maximum at a lower energy $\delta E \approx
100\,\mu$eV. This might seem surprising at first glance as we expect to inject
electrons only at one energy; but we can understand this by taking into account
tunnelling through excited states of the detector dot which occurs at lower
values of $\Delta E$ as depicted in Fig.~\ref{fig3}a. Each excited state
contributes to the measured spectrum a replica of the energy distribution of
the impinging electrons at apparently lower energy . Hence we measure a
convolution of the true signal with the response function of the detector,
which we can model as a sum of weighted $\delta$-functions. The different
excitation spectra of QD1 and QD2 are responsible for the differences in the
energy dependence of the measured currents in Fig.~\ref{fig2}a and \ref{fig2}b.

The excitation energies of the dot and the respective tunnelling rates can be
determined from transport spectroscopy of the detector quantum dot. The
measurement shown in Fig.~\ref{fig3}b shows not only the ground state resonance
with the source chemical potential (marked by a white line) but also resonances
due to a strong excited state with excitation energy $\varepsilon_1 =
90\,\mu$eV (solid red line) and a further weak excited state at $\varepsilon_2=
160\,\mu$eV (dotted red line, revealed by a negative $dI/d\Vsd$). We can
estimate the ratio of the tunnelling rates as $r_1=\Gamma_1/\Gamma_0 \approx
2.5$ and $r_2=\Gamma_2/\Gamma_0 \approx 1$. We can now use these results to
perform a deconvolution of the measured spectrum as follows: the desired signal
which would be measured in the absence of excited states is represented by a
equi-spaced cubic spline.  The convolution of this trial function with the
response function
$\delta(E)+r_1\delta(E-\varepsilon_1)+r_2\delta(E-\varepsilon_2)$ is
calculated, and a linear least-squares fit to the measured current then
determines the spline coefficients and thus the spectrum. Fig.~\ref{fig3}c
shows the normalized detector current for a measurement with injection at
$-e\Vsd = 230\,\mu$eV. The result of the deconvolution is displayed in
Fig.~\ref{fig3}d. The replica signals at lower energy were largely removed and
the true spectrum of ballistic electrons with a narrow energy distribution is
revealed.

\begin{figure}[tb]
    \begin{center}
            \includegraphics[width=\plotwidth]{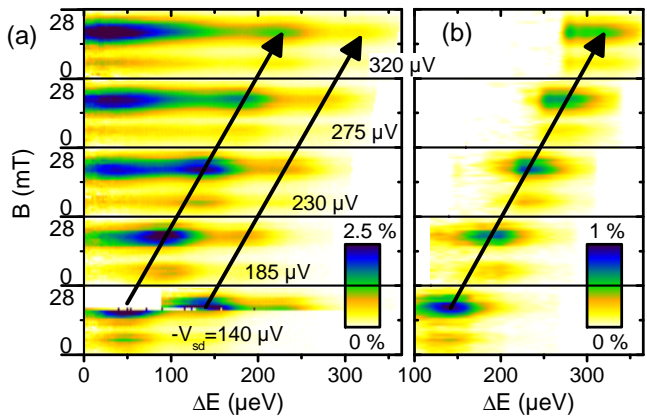}
    \end{center}
        \vspace{-4mm}
    \caption{%
a) Normalized detector signal for five different injector bias voltages.
%(injection by QD1, detection by QD2).
b) Result of the deconvolution taking
into account two excited states. The arrows indicate the expected energy shift
of the ballistic signal.
        }
            \vspace{-2mm}
    \label{fig4}
\end{figure}

Now we have demonstrated ballistic electron spectroscopy for a single injection
energy, we proceed by varying the excess energy. Fig.~\ref{fig4} shows the
spectra resulting from varying $\delta E = -e\Vsd$. For each $\Vsd$ value of
the injector dot, its plunger gate voltage $\Vp$ was adjusted to achieve
resonance of the injector level with the source chemical potential
(Fig.~\ref{fig1}). It is clear that the raw measured signal in Fig.~\ref{fig4}a
demonstrates the expected shift of the spectrum with increasing injection
energy. This becomes even clearer in the deconvolved signal of
Fig.~\ref{fig4}b. Here we used the same tunnelling rate ratios $r_{1,2}$ as in
Fig.~\ref{fig3}d. Our simplified assumption of constant tunnelling rates
accounts for the reduced quality of the deconvolution at high energies; a
closer look to Fig.~\ref{fig3}b reveals a voltage dependence of the conductance
along the marked lines and thus a change of tunnelling rates. A more elaborate
characterization of the detector dot would allow an improved quality of
deconvolution.

Fig.~\ref{fig4} also reveals that the amplitude of the spectrometer signal at
$\Delta E = \delta E = -e\Vsd$ falls with rising energy while the low energy
signal near $\Delta E = 0$ increases. At the highest bias the reduction amounts
to about $50\,$\%. If we assume scattering as the only cause of signal
reduction we deduce a scattering length of $l_{s}\sim 4\,\mu$m and a scattering
time of about $\tau \sim 25\,$ps. The predicted value for pure
electron-electron scattering for excess energy $\delta E = 320\,\mu$eV and
Fermi energy $E_F = 6\,$meV is $l_{e-e} \sim
10\,\mu$m.\cite{Giuliani1982,Yacoby1991} Further experiments are needed to test
whether the smaller experimental value is caused by additional scattering or
some other effect.

%%%%%%%%%%%%%%%%%%%%%%%%%%%%%%%%%%%%%%%%%%%%%%%%%%%%%%%%%%%%%%%%%%%%%
%%%%%    Conclusion    %%%%%%%%%%%%%%%%%%%%%%%%%%%%%%%%%%%%%%%%%%%%%
%%%%%%%%%%%%%%%%%%%%%%%%%%%%%%%%%%%%%%%%%%%%%%%%%%%%%%%%%%%%%%%%%%%%%

In conclusion, we have demonstrated the use of a quantum dot as a ballistic
electron spectrometer. We have tested the scheme using a second quantum dot to
source non-equilibrium ballistic electrons with a narrow energy and angle
distribution. Using deconvolution techniques we can extract the spectrum even
in the presence of excited states of the detector dot. The performance will
increase further if we engineer the detector dot for larger excitation energy
of a few $100~\mu$eV.

Among the many potential applications of ballistic electron spectroscopy we can
mention only a few: For quantum wires the position and movement of the subband
energies could be measured near the 0.7 anomaly; the enhanced density of states
of a quantum dot in the Kondo regime could be traced; resonant tunnelling
involving non-equilibrium dot states could be characterized and the energy
relaxation of non-equilibrium electrons could be measured in detail.

%%%%%%%%%%%%%%%%%%%%%%%%%%%%%%%%%%%%%%%%%%%%%%%%%%%%%%%%%%%%%%%%%%%%%
%%%%%    Acknowledgement    %%%%%%%%%%%%%%%%%%%%%%%%%%%%%%%%%%%%%%%%%%%%%
%%%%%%%%%%%%%%%%%%%%%%%%%%%%%%%%%%%%%%%%%%%%%%%%%%%%%%%%%%%%%%%%%%%%%
The authors would like to thank Abi~Graham, Chris~Ford and David~Anderson. We
acknowledge financial support from EPSRC and the European Union within the
research training network COLLECT.

%%%%%%%%%%%%%%%%%%%%%%%%%%%%%%%%%%%%%%%%%%%%%%%%%%%%%%%%%%%%%%%%%%%%%
%%%%%    References  %%%%%%%%%%%%%%%%%%%%%%%%%%%%%%%%%%%%%%%%%%%%%%%%%%
%%%%%%%%%%%%%%%%%%%%%%%%%%%%%%%%%%%%%%%%%%%%%%%%%%%%%%%%%%%%%%%%%%%%%

%\bibliographystyle{apsrev}
%\bibliography{../bib/spectroscopy-references}

\end{document}